\def\year{2018}\relax
\documentclass[letterpaper]{article} 
\usepackage{aaai18}  
\usepackage{times}  
\usepackage{helvet}  
\usepackage{courier}  
\usepackage{url}  
\usepackage{graphicx}  
\usepackage[printwatermark]{xwatermark}
\usepackage{xcolor}
\usepackage{balance}
\usepackage{booktabs} 
\usepackage{multirow}

\usepackage{amsmath}
\usepackage{amssymb}
\DeclareMathAlphabet{\pazocal}{OMS}{zplm}{m}{n}
\DeclareMathAlphabet{\pazocal}{OMS}{zplm}{m}{n}

\newcommand{\Ds}{\pazocal{D}}
\newcommand{\Vs}{\pazocal{V}}

\newcommand{\Ls}{\pazocal{L}}

\DeclareMathOperator \real{\mathbb{R}}

\begin{document}
%

\title{Graph Based Semi-supervised Learning with Convolution Neural Networks \\to Classify Crisis Related Tweets}
 \author{Firoj Alam\textsuperscript{1}, Shafiq Joty\textsuperscript{2}, Muhammad Imran\textsuperscript{1}\\
  \textsuperscript{1}{Qatar Computing Research Institute, HBKU, Doha, Qatar}\\
  \textsuperscript{2}{School of Computer Science and Engineering, NTU, Singapore}\\
  \textsuperscript{1}\{fialam, mimran\}@hbku.edu.qa,
  \textsuperscript{2}srjoty@ntu.edu.sg
}
  

\maketitle
\begin{abstract}
During time-critical situations such as natural disasters, rapid classification of data posted on social networks by affected people is useful for humanitarian organizations to gain situational awareness and to plan response efforts. However, the scarcity of labeled data in the early hours of a crisis hinders machine learning tasks thus delays crisis response. In this work, we propose to use an inductive semi-supervised technique to utilize unlabeled data, which is often abundant at the onset of a crisis event, along with fewer labeled data. Specifically, we adopt a graph-based deep learning framework to learn an inductive semi-supervised model. We use two real-world crisis datasets from Twitter to evaluate the proposed approach. Our results show significant improvements using unlabeled data as compared to only using labeled data.
\end{abstract}

\section{Introduction}
The recent emergence and wide-adaptation of microblogging platforms such as Twitter, during crises and emergency situations due to natural or man-made disasters, has been proven useful for a number of humanitarian tasks~\cite{imran2015processing}. Affected population post timely and useful information of various types such as reports of injured or dead people, infrastructure damage, urgent needs (food, shelter, medical assistance) on these social networks. Humanitarian organizations believe timely access to this important information 
in the first few hours 
can help significantly and can reduce both human loss and economic damage~\cite{vieweg2014integrating,varga2013aid}.

In order to identify useful messages for humanitarian tasks one potential approach is to use supervised learning to automatically categorize each incoming message (e.g., tweets) into one of the two classes i.e., \textit{relevant} and \textit{irrelevant} \cite{ICWSM1715655}. In order to design the classification model, obtaining a large amount of labeled data is a challenging task, particularly during the first few hours of a crisis situation. However, access to abundant unlabeled data is possible under such time-critical situations, as hundreds of tweets arrive each minute. Moreover, one can rely on labeled data from past similar events. 
In such situations, semi-supervised methods can provide effective ways to leverage unlabeled data in addition to labeled data. 

Many models have been proposed for semi-supervised learning including generative models \cite{nigam2000text}, co-training, \cite{mitchell1999role}, 
self-training \cite{mihalcea2004co}, and graph-based models \cite{subramanya2014graph}. These methods can be categorized into two types: \emph{transductive} and \emph{inductive}. In the transductive setting, a learner is only applicable to the unlabeled instances observed at training time, that is, the learner does not generalize to unobserved instances. Whereas, an inductive learner generalizes to data that are not seen at the training time. Therefore, it is more desirable to have an inductive learner over a transductive one. Other reason to prefer inductive semi-supervised learning over the transductive approach is that it avoids building the graph each time it needs to infer the labels for the unlabeled instances. 

Among other semi-supervised text classification approaches, Johnson et al. \cite{johnson2015semi} use a Convolutional Neural Networks (CNN) via region embedding in which the CNN learns a small region from embedding. Miyato et al.~\cite{miyato2016adversarial} used adversarial training for text classification with a small perturbations on the input word embeddings. 
Our motivation of using deep neural network (i.e., CNN) is that it has shown a great success in recent years in many different areas such as NLP and data mining \cite{collobert2011natural,Grover.Leskovec:16}. Apart from the improved performance, one crucial benefit of DNN is that they obviate the need for feature engineering and learn latent features automatically as distributed dense vectors. This capability of DNN has recently been extended to the semi-supervised setting \cite{yang2016revisiting}.

In this work, we adopt a graph-based deep learning framework recently proposed by Yang et. al \cite{yang2016revisiting} for learning an inductive semi-supervised model to classify tweets in a crisis situation. 
In this framework, CNN is combined with graph-based network that learns internal representations of the input by predicting contextual nodes in a graph that encodes \emph{similarity} between labeled and unlabeled training instances.
Compared to the work of Yang et. al \cite{yang2016revisiting}, our approach is different in several ways: 1) we construct a graph by computing the distance between tweets based on word embeddings, 2) we use a CNN to compose higher-level features from the word embeddings,
and 3) for context prediction, instead of performing a random walk, we select nodes based on their similarity in the graph.  

The evaluation of the proposed approach (see Sec. Methodology) is conducted using two real-world Twitter datasets. Our results (see Sec. Results and Discussion) demonstrate the effectiveness of our approach with an improvement from $5\%$ to $26\%$ compared to the supervised classification. The experimental data can be accessed through \url{http://crisisnlp.qcri.org}.

\section{Methodology}
\label{sec:methodology}

Let $\Ds_l = \{\mathbf{t}_i, y_{i}\}_{i=1}^{L}$ and $\Ds_u = \{\mathbf{t}_i\}_{i=1}^{U}$ be the set of labeled and unlabeled tweets for a particular crisis event, where $y_i \in \{1, \ldots, K\}$ is the class label for tweet $\mathbf{t}_i$, $L$ and $U$ are the number of labeled and unlabeled tweets, respectively, with $n=L+U$ being the total number of training instances. Our goal is to learn an inductive model $p(y_i|\mathbf{t}_i, \theta)$, where $\theta$ denotes the model parameters. We adopt the graph-based semi-supervised embedding learning framework proposed in \cite{yang2016revisiting}. In this framework, in addition to the labeled data, a ``similarity'' graph $G$ is used to learn internal representations for the input (i.e., embeddings) by exploiting relations between labeled and unlabeled training instances. 

\subsection{Graph Construction}
\label{ssec:graph-consturction}
Given a set of $n$ instances (tweets in our case), a typical approach is to construct the graph based on relational knowledge source (e.g., citation links in \cite{lu:icml03}) or distance between instances \cite{zhu05survey}. 
However, developing such a relational knowledge is not feasible for every problem.
On the other hand, computing distance between $n(n-1)/2$ pairs of instances to construct the graph is also very expensive \cite{muja2014scalable}. Hence, we choose to use k-nearest neighbor-based approach for finding nearest neighbors of instances as it has been shown to be an effective approach in other studies \cite{dong2011efficient,jebara2009graph}. 
The nearest neighbor graph consists of $n$ vertices and for each vertex, there is an edge set consisting of a subset of $n$ instances. The edge is defined by the distance measure $d({i},{j})$ between tweets $\mathbf{t}_{i}$ and $\mathbf{t}_{j}$, where the value of $d$ represents how similar the two tweets are. Similar similarity-based graph has shown impressive results in learning sentence representations \cite{saha-17}. 
To find the nearest instances efficiently, we used k-d tree data structure \cite{witten2016data}. 
The rationale of this approach is that if $\mathbf{t}_{i}$ is very far from $\mathbf{t}_{j}$ and $\mathbf{t}_{k}$ is close to $\mathbf{t}_{j}$ then without computing the distance between $\mathbf{t}_{i}$ and $\mathbf{t}_{k}$ we can infer they are far. 
For our graph construction, we first represent each tweet by averaging the word embedding vectors (see Sec. Crisis Word Embedding for details) of its words, and then we measure $d(i,j)$ by computing the \textit{Euclidean} distance between the vectors. We have chosen to use \textit{Euclidean} distance to reduce computational complexity. The number of nearest neighbor $k$ was set to 10.

\subsection{Semi-supervised Neural Network Model}
\label{ssec:semi-supervised}

\begin{figure}[t]
	\includegraphics[height=2.3in]{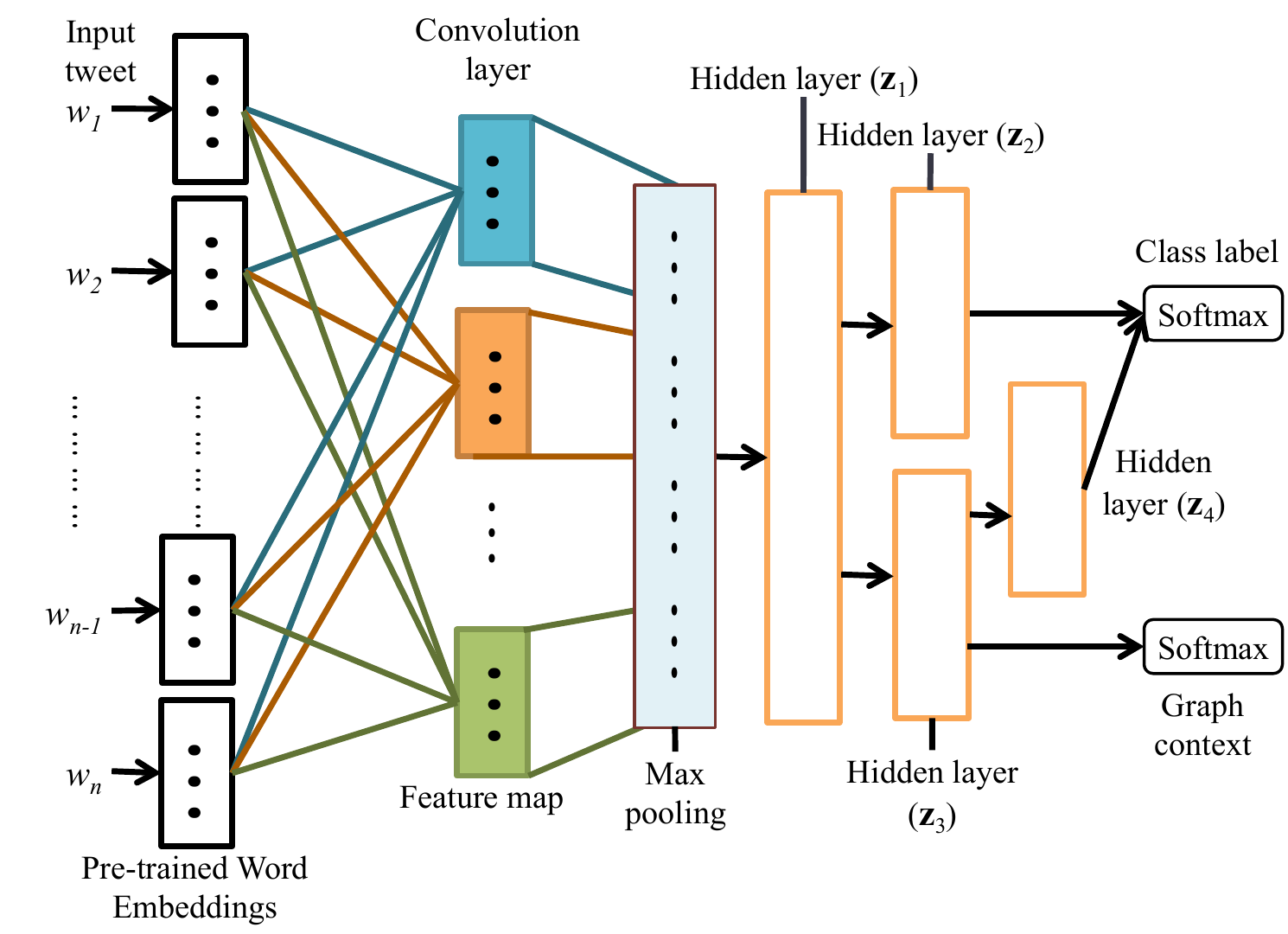}
	\caption{The architecture of the graph-based semi-supervised learning with CNN.}
\label{fig:cnn_graph}
\end{figure}

Figure \ref{fig:cnn_graph} shows the architecture of our neural network model. The input to the network is a tweet $\mathbf{t}=(w_1, \ldots, w_n)$ containing words each coming from a finite vocabulary $\Vs$. The first layer of our network maps each of these words into a distributed representation $\real^d$ by looking up a shared embedding matrix $E$ $\in$ $\real^{|\Vs| \times d}$. We initialized $E$ using pretrained word vectors (more in Sec. Crisis Word Embedding). The output of the look-up layer is a matrix  $X \in \real^{n \times d}$, which is passed through a number of convolution and pooling layers to learn higher-level feature representations. 

A convolution operation applies a \emph{filter} $\mathbf{u} \in \real^{k.d}$ to a window of $k$ vectors to produce a new feature $h_t = f(\mathbf{u} . X_{t:t+k-1})$, where $X_{t:t+k-1}$ is the concatenation of $k$ look-up vectors, and $f$ is a nonlinear activation; we use rectified linear units (ReLU). We apply this filter to each possible $k$-length windows in $X$ to generate a \emph{feature map}, $\mathbf{h}^j = [h_1, \ldots, h_{n+k-1}]$. This  process is repeated $N$ times with $N$ different filters to get $N$ different feature maps. We then apply a \emph{max-pooling} operation, $\mathbf{m} = [\mu_p(\mathbf{h}^1), \cdots, \mu_p(\mathbf{h}^N)]$, where $\mu_p(\mathbf{h}^j)$ refers to the $\max$ operation applied to each window of $p$ features in the feature map $\mathbf{h}^i$. Intuitively, the filters compose local features into higher-level representations in the feature maps, and max-pooling extracts the most important aspects from each feature map while reducing the output dimensionality. 
The pooled features are passed through two fully-connected hidden layers, $\mathbf{z}_1 = f(V_1\mathbf{m})$; $\mathbf{z}_2 = f(V_2\mathbf{z}_1)$, where $V_1$ and $V_2$  are the associated weight matrices. The final activations are used for classification using a Softmax in the output layer; the formal definition of Softmax is defined below. 

To leverage the unlabeled data $\Ds_u$, and to exploit the relations between training instances (labeled or unlabeled) encoded in the graph $G$ we add a branch to the network. It takes $\mathbf{z}_1$ as input and learns internal representations by predicting a node in the graph context of the input tweet. Following \cite{yang2016revisiting}, we use \emph{negative sampling} to compute the loss for predicting the  context node, and we sample two types of contextual nodes: one is based on the graph $G$ to encode the structural information and the second is based on labels to incorporate label information through this branch of the network. The ratio of positive and negative samples is controlled by a random variable $\rho_1 \in (0,1)$, and the proportion of the two context types is controlled by another random variable $\rho_2 \in (0,1)$; see Alg. 1 of \cite{yang2016revisiting} for details of the sampling procedure. Let $(i, j, \gamma)$ is a tuple sampled from the distribution $p(i, j, \gamma|\Ds_l, \Ds_u, G)$, where $j$ is a context node of an input node $i$ and $\gamma \in \{+1,-1\}$ denotes whether it is a positive or a negative sample; $\gamma = +1$ if $\mathbf{t}_i$ and $\mathbf{t}_j$ are neighbors in the graph (for graph-based context) or they both have same labels (for label-based context), otherwise $\gamma = -1$. The loss for context prediction can be written as $\Ls_c (\theta) = \mathbb{E}_{(i, j,\gamma)} \log \sigma \left( \gamma \mathbf{w}_{j}^{T} \mathbf{z}_3 (i) \right)$, where $\mathbf{z}_3 (i) = f(V_3 \mathbf{z}_1 (i))$ defines another fully-connected hidden layer (marked as \emph{Hidden layer ($\mathbf{z}_3$)} in Fig. \ref{fig:cnn_graph}) having weights $V_3$, and $\mathbf{w}_j$ is the weight vector associated with the context node $\mathbf{t}_j$.

For the classification, we take $\mathbf{z}_3$ and pass it through another fully-connected hidden layer, $\mathbf{z}_4 = f(V_4 \mathbf{z}_3)$, where $V_4$ is the corresponding weight matrix. Finally, the Softmax output layer does the classification  $p(y=k|\mathbf{t}, \theta) = {\exp \left( \mathbf{w}_{k}^T \left[\mathbf{z}_2; \mathbf{z}_4 \right] \right)}/{\sum_{k'} \exp \left( \mathbf{w}_{k'}^T \left[\mathbf{z}_2; \mathbf{z}_4 \right] \right)}$, where $[.;.]$ denotes concatenation of two column vectors, and $\mathbf{w}_{k}$ are the class weights. The overall loss of the network can be written as  $\Ls (\theta) = {- \frac{1}{L} \sum_{i=1}^{L} \log p(y_{i}|\mathbf{t}_{i},\theta)} - {\lambda \Ls_c(\theta)}$, where the first part is the classification loss and the second part is the context loss, and the hyperparameter $\lambda$ controls the relative strength of the two parts.

\subsection{Crisis Word Embedding}
\label{ssec:word_embedding}
We initialize the embedding matrix $E$ in our network with pretrained word embeddings. We trained a continuous bag-of-words (CBOW) wrod2vec \cite{mikolov2013efficient} model on a large crisis dataset with vector dimensions of $300$, a context window size of 5 and $k=5$ negative samples. The crisis dataset consists of different collections of tweets collected automatically using the AIDR system \cite{imran2014aidr}. In the preprocessing, we lowercased the tweets and removed URLs, digit, time patterns, special characters, single character, user name started with the $@$ symbol. The resulting dataset has about 364 million tweets and about 3 billion words. When training CBOW, we filtered out words with a frequency less than or equal to 5. The resulting trained word-embedding model contains about 2 million words.

\section{Experiments}
\label{sec:experiments}
In this section, we first describe the datasets we used in our experiments, then the experimental setting, and finally the results. 

\subsection{Datasets}
\label{ssec:data}
For the evaluation, we use two real-world Twitter datasets collected during the \textit{2015 Nepal earthquake} and the \textit{2013 Queensland floods}. These datasets are 
collected through the Twitter streaming API\footnote{https://dev.twitter.com/streaming/overview} using event-specific keywords/hashtags. To obtain the labeled examples for our task which consists of two classes \textit{relevant} and \textit{irrelevant}, we employed paid workers from the Crowdflower\footnote{http://crowdflower.com} -- a crowdsourcing platform. We randomly sampled 11,668 and 10,033 tweets from the Nepal earthquake and the Queensland floods respectively. Given a tweet, we asked crowdsourcing workers to assign the \textit{``relevant''} label if the tweet conveys/reports information useful for crisis response such as a report of injured or dead people, some kind of infrastructure damage, urgent needs of affected people, donations requests or offers, otherwise assign the \textit{``irrelevant''} label.

For evaluation, we split the datasets into 60\% as training, 30\% as test and 10\%  as development. Table~\ref{table:data_dist} shows the resulting datasets.

\begin{table}[t]
\centering
\caption{Distribution of the labeled datasets}
\label{table:data_dist}
\scalebox{0.9}{
\begin{tabular}{|l|r|r|r|r|r|}
\hline
\multicolumn{1}{|c|}{\textbf{Data}} & \multicolumn{1}{c|}{\textbf{Relevant}} & \multicolumn{1}{c|}{\textbf{Irrelevant}} &\multicolumn{1}{c|}{\textbf{Train}} & \multicolumn{1}{c|}{\textbf{Dev}} & \multicolumn{1}{c|}{\textbf{Test}} \\ \hline
Nepal & 5,527 & 6,141 & 7,000 & 1,166 & 3,502 \\ \hline
Queensland & 5,414 & 4,619 &6,019 & 1,003 & 3,011 \\ \hline
\end{tabular}
}
\vspace{-1em}
\end{table}

\subsection{Experimental Settings}
\label{ssec:exp_settings}
As a part of the preprocessing of the dataset, we used the same approach that we used to train the word2vec model (see Sec. Crisis Word Embedding). We use the validation set to optimize the hyperparameters. 

For our semi-supervised setting, one of the main goals was to understand how much labeled data is sufficient to obtain a reasonable result. Therefore, we experimented our system considering the smallest to all instances, such as 100, 500, 2000, 5000 and all instances. Such an understanding can help us to design the model at the onset of a crisis event with sufficient amount of labeled data. To demonstrate that the semi-supervised approach outperforms the supervised baseline, we run supervised experiments using the same number of labeled instances. In the supervised setting, only $\mathbf{z}_2$ activations in Fig. \ref{fig:cnn_graph} are used for classification.  

We trained the models using the adadelta \cite{zeiler2012adadelta} algorithm. The learning rate was set to $0.1$ when optimizing on the classification loss and to $0.001$ when optimizing on the context loss. The maximum number of epochs was set to 200, and dropout \cite{srivastava2014dropout} rate of $0.02$ was used to avoid overfitting. We did \emph{early stopping} based on the f-measure on the validation set with a patience of 25.

We used 100, 150 and 200 filters each having the window size of 2, 3 and 4, respectively, and pooling length of 2, 3 and 4 respectively. The value of $\lambda$ was set to $1.0$ in the semi-supervised model. We did not tune any hyperparameter (e.g., the size of hidden layers, filter size, dropout rate) in any experimental setting since the goal was to have an end-to-end comparison for the same hyperparameter setting and understand whether CNN with graph based semi-supervised approach can outperform or not. We did not filter out any vocabulary item for any of the settings. We also did not fine-tune the word embeddings on the classification task.

\subsection{Results and Discussion}
\label{ssec:results_and_discussion}

\begin{table}[h]
\centering
\caption{F-measure for different experimental settings. \textit{L} refers to labeled data, \textit{U} refers to unlabeled data, \textit{All L} refers to all labeled instances for that particular dataset.}
\label{table:classification_results}
\scalebox{0.75}{
\begin{tabular}{|l|l|r|r|r|r|r|}
\hline
\multicolumn{1}{|c|}{\textbf{Dataset}} & \multicolumn{1}{c|}{\textbf{L/U}} & \multicolumn{1}{c|}{\textbf{100}} & \multicolumn{1}{c|}{\textbf{500}} & \multicolumn{1}{c|}{\textbf{1000}} & \multicolumn{1}{c|}{\textbf{2000}} & \multicolumn{1}{c|}{\textbf{All L}} \\ \hline
\multirow{2}{*}{\textbf{Nepal Earthquake}} & L & 47.11 & 52.63 & 55.95 & 58.26 & 60.89 \\ \cline{2-7} 
 & L+U(50k) & 52.32 & 59.95 & 61.89 & 64.05 & 66.63 \\ \hline
\multirow{2}{*}{\textbf{Queensland Flood}} & L & 58.52 & 60.14 & 62.22 & 73.92 & 78.92 \\ \cline{2-7} 
 & L+U($\sim$21k) & 75.08 & 85.54 & 89.08 & 91.54 & 93.54 \\ \hline
\end{tabular}
}
\vspace{-1em}
\end{table}

In Table \ref{table:classification_results}, we present the classification results for different experimental settings. We computed the performances using weighted averaged precision, recall and F-measure. In the table, we only report the F-measure for simplicity. The rational behind choosing the weighted metric is that it takes into account the class imbalance problem. From the table, we see that as we increase the number of labeled examples ($\textbf{L}$), the classification performance improves -- from $47.1$ to $60.9$ for Earthquake and from $58.5$ to $78.9$ for Flood, which is a common trend for supervised models. 

In this study, for computational efficiency, we limit the number of unlabeled instances $\textbf{U}$ in our semi-supervised model to $50K$ for Nepal and $\sim$21K for Queensland. We can observe that as we include unlabeled instances with labeled instances, performance significantly improves in each experimental setting giving $5\%$ to $26\%$ absolute improvements over the supervised models. These improvements demonstrate the effectiveness of our approach. We also notice that our semi-supervised approach can perform above $90\%$ depending on the event. 
Recall that we did not tune the hyperparameters of our supervised and semi-supervised models. Therefore, these results may not be optimal. We believe that upon optimizing hyperparameters, the overall performances of our system can be further improved. 
From the results, we can ascertain that 500 labeled instances with unlabelled instances could be a reasonable choice at the early onset of a crisis event to design the semi-supervised model. 

\section{Conclusions}
\label{sec:conclusions}
We presented a graph-based semi-supervised deep learning framework based on a CNN. The network combines a loss for predicting the class labels with a loss for predicting the context defined by a similarity graph. We constructed the similarity graph using a k-nearest neighbor approach that exploits distributed representations of tweets. Our evaluation on two crisis-related tweet datasets demonstrates significant improvements for our semi-supervised model over a supervised only baseline. There are several interesting future research directions of this work such as exploring domain adaptation and zero- or one-shot learning. 


{
\footnotesize
\bibliographystyle{aaai}
\balance
\bibliography{./bib/all}
}
\end{document}